# Ferroelectric Chirality-Driven Direction-Tunable and Spin-Invertible Corner States in 2D MOF-Based Magnetic Second-Order Topological Insulators

*Jialin Gong, Wei Sun, Yang Wu, Zhenzhou Guo, Shifeng Qian\*, Xiaotian Wang\*, Gang Zhang\**

J. Gong, Z. Guo, X. Wang

Institute for Superconducting and Electronic Materials, Faculty of Engineering and Information Sciences, University of Wollongong, Wollongong, New South Wales 2500, Australia.
*E-mail: xiaotianw@uow.edu.au

W. Sun

Shandong Provincial Key Laboratory of Preparation and Measurement of Building Materials, University of Jinan, Jinan, 250022, China

Y. Wu

Bremen Center for Computational Materials Science, University of Bremen, Bremen 28359, Germany

S. Qian

Anhui Province Key Laboratory for Control and Applications of Optoelectronic Information Materials, Department of Physics, Anhui Normal University, Wuhu, Anhui 241000, China
*E-mail: qiansf@ahnu.edu.cn

G. Zhang

Yangtze Delta Region Academy of Beijing Institute of Technology, Jiaxing 314000, China
*E-mail: zhanggang@bitjx.edu.cn

Despite the rapid progress in predicting 2D magnetic second-order topological insulators (SOTIs), effective strategies for manipulating their spin-polarized corner states remain largely unexplored. The interplay between ferroelectricity, chirality, magnetism, and topology presents an untapped opportunity for controlling these corner states. Here, we propose a novel approach for tuning spin-polarized corner states in 2D magnetic SOTIs by inducing ferroelectric chirality in 2D metal-organic frameworks (MOFs) with intrinsic structural flexibility. Through symmetry analysis, we strategically replace pyrazine (pyz) ligands with 2-pyrazinolate (2-pyzol) ligands in the 2D MOF Cr(pyz)$_2$, leading to the emergence of a new 2D magnetic SOTI, Cr(2-pyzol)$_2$, which facilitates ferroelectric chirality controlled spin-polarized corner states in both spin channels. Through first-principle calculations, we demonstrate that Cr(2-pyzol)$_2$ belongs to ferroelectric chiral systems, and its corner states can be directionally tuned in real space and spin-inverted in spin space upon ferroelectric chirality switching. Our work represents the first attempt to simultaneously manipulate corner states in both real space and spin space, offering a new strategy for integrating ferroelectric chirality into 2D MOF-based magnetic SOTIs.

# 1. Introduction

Chirality[1-3] is a fundamental property of nature, describing objects that cannot be superimposed on their mirror images—analogous to the relationship between left and right hands. It is a key research frontier[4-8] spanning materials science, physics, chemistry, and biology. Recent advancements in field-tuned topological formations in nanostructured ferroelectrics have introduced controllable topological chirality[9-13], opening new opportunities for developing *ferroelectric chiral materials* with practical applications[14-20]. The interplay between ferroelectric polarization and chirality in such materials gives rise to key phenomena, including: (i) ferroelectric polarization reversal inducing chirality switching, enabling multifunctional applications, and (ii) intrinsic chirality modulating the direction of ferroelectric polarization, thereby affecting the ferroelectric response.

Meanwhile, the discovery of topological insulators (TIs)[21-23] has significantly impacted the fields of condensed matter physics and materials science. The conventional bulk-boundary correspondence is a fundamental property of TIs, namely, a first-order TI with a dimension of $d$ hosts boundary states with a dimension of $(d-1)$. In recent years, the concept of bulk-boundary correspondence has been expanded to include unconventional cases, resulting in higher-order ($n$th-order and $n>1$) TIs[24-30] with a dimension of $d$ that exhibit $(d-n)$-dimensional boundary states. For example, a 2D second-order TI (SOTI) demonstrates protected boundary states that are localized at 0D corners. Beyond nonmagnetic 2D SOTIs, a growing research focus is on integrating magnetism with second-order topology, leading to the discovery of numerous 2D magnetic SOTI candidates[31-40]. These candidates host unconventional boundary states, i.e., spin-polarized corner states, advancing the study of spin-polarized topological quantum materials. However, most reported magnetic SOTIs[32,34-37,39,40] have inversion symmetry (***I***), which prevents their classification as ferroelectric chiral materials. As a result, ferroelectric chirality cannot control the corner states in these magnetic SOTIs that are protected by ***I***.

This raises key open questions: *Can ferroelectric chirality manifest in 2D magnetic SOTIs? If so, can it be used to regulate spin-polarized corner states?* To our knowledge, these questions remain unanswered. A major challenge lies in the scarcity of suitable materials that simultaneously exhibit magnetism, second-order topology, chirality, and ferroelectricity. To address this, in this work, we focus on 2D MOFs as potential candidates due to the following advantages: (i) MOFs offer a promising platform for 2D ferroelectric chiral materials[41-49]. The main strategies for achieving chirality in MOFs involve the induction of chirality in achiral frameworks via external chiral agents and the direct incorporation of chiral linkers, chiral nodes, or both. (ii) The chemical flexibility in selecting metal nodes, organic linkers, and their connectivity enables the easy tuning of the electronic and magnetic properties of 2D MOFs, making them ideal candidates for the development of magnetic topological states [37,50-55].

In this work, we employ the 2D MOF Cr(pyz)$_2$ as a representative material and aim to provide a strategy to realize ferroelectric chirality and tunable spin-polarized corner states by multiple symmetry breaking. Cr(pyz)$_2$, constructed through Cr nodes connected by pyz as organic linkers, has been reported as a magnetic SOTI [51] and can be exfoliated from the experimentally synthesized layered crystal Li$_{0.7}$[Cr(pyz)$_2$]Cl$_{0.7}$·0.25(THF) (where THF refers to tetrahydrofuran)[56]. Symmetry analysis indicates that maintaining out-of-plane rotational symmetry is essential for preserving a quantized fractional corner charge while breaking both the $M_z$ mirror symmetry and in-plane $C_2$ rotational symmetry is necessary for sustaining non-vanishing z-direction polarization. Leveraging the inherent flexibility of MOF Cr(pyz)$_2$, we achieve these symmetries breaking via a simple modification: replacing four of the sixteen hydrogen atoms in pyz ligands with oxygen to form 2-pyzol ligands. This substitution transforms Cr(pyz)$_2$ into Cr(2-pyzol)$_2$, which exhibits right-handed chirality ($C_R$) and electric polarization inward perpendicular to the a-b lattice plane ($-P_z$), belonging to the space group P2 (No. 3). Using first-principles calculations, we demonstrate that Cr(2-pyzol)$_2$ exhibits quantized fractional topological charge, ferroelectric chirality, and a magnetic SOTI phase. Furthermore, we show that its corner states are directionally tunable and spin-invertible upon ferroelectric chirality switching. Additional computational details are provided in the Supporting Information (SI).

## 2. Results and Discussion

### 2.1. Symmetry analysis

We first analyze the symmetry requirements for manipulating spin-polarized corner states through ferroelectric chirality control. Since our objective is to realize a magnetic ferroelectric phase, both time-reversal symmetry ($T$) and $I$ must be fundamentally broken. However, achieving ferroelectric chirality not only necessitates breaking $I$ but also requires careful consideration of symmetry constraints. For instance, the combination of $C_{2z}$ rotational and $M_z$ mirror symmetries would result in vanishing polarization. Meanwhile, maintaining both fractional quantized corner charge and non-zero polarization along the z-direction, we must preserve out-of-plane rotational symmetry while breaking both in-plane $C_2$ rotational and $M_z$ mirror symmetries under the prerequisite of $I$ and $T$ symmetries breaking. This specific symmetry configuration enables the coexistence of z-directional polarization and quantized fractional corner charge. Furthermore, to achieve tunable corner states, we propose that lower rotational symmetry constraints would be preferable, with $C_{2z}$ symmetry emerging as the optimal choice for system design.

The Cr(pyz)$_2$ MOF consists of planar tetracoordinated Cr atoms bonded to four neighboring pyz ligands, forming a 2D orthorhombic lattice. As shown in Figure 1, the crystal structure of Cr(pyz)$_2$ displays $I$, $C_{4z}$, and $M_z$ symmetries and is classified under space group P4/nbm (No. 125). The presence of $I$ symmetry prohibits ferroelectric chirality, while $C_{4z}$ symmetry forces corner states to localize at the four corners of a square-shaped nanodisk, leading to direction-independent behavior (see Figure S1 and Table S1 of the SI, and Ref. [51]). Although Cr(pyz)$_2$ itself lacks the desired properties, the inherent structural flexibility of MOFs provides a promising platform for achieving targeted characteristics through strategic structural modifications.

To induce ferroelectric chirality and tunable corner states in Cr(pyz)$_2$, multiple symmetry breaking should be considered. The parent structure enforces vanishing polarization due to $I$ symmetry, necessitating its breaking. Furthermore, the combination of $C_{4z}$ and $M_z$ operators also suppresses polarization in Cr(pyz)$_2$. If in-plane polarization is to be generated, both $C_{4z}$ and $C_{2z}$ symmetries must be broken, whereas, for quantized fractional corner charges, rotational symmetry along the z-axis must be retained. We therefore break $M_z$ mirror symmetry to induce out-of-plane polarization while retaining partial rotational symmetry. Additionally, we further break $C_{4z}$ symmetry while preserving $C_{2z}$ symmetry, enabling switchable corner states. This specific symmetry manipulation also allows direction-dependent corner states to interact with different spins, thus facilitating electric-field control over both topological states and magnetism.

As shown in Figure 1, one can achieve these symmetries breaking in MOF Cr(pyz)$_2$ through a simple structural modification—replacing four (out of 16) hydrogen atoms with oxygen (i.e., substituting pyz with 2-pyzol) in MOF Cr(pyz)$_2$. As a result, MOF Cr(2-pyzol)$_2$ with right-handed chirality ($C_R$) was obtained, lacking $I$, and exhibiting $-P_z$ and chirality ($C$), along with direction-dependent corner states. Interestingly, the corner states can be directionally tuned in real space and spin-inverted in spin space when ferroelectric chirality is switched, as will be discussed further.

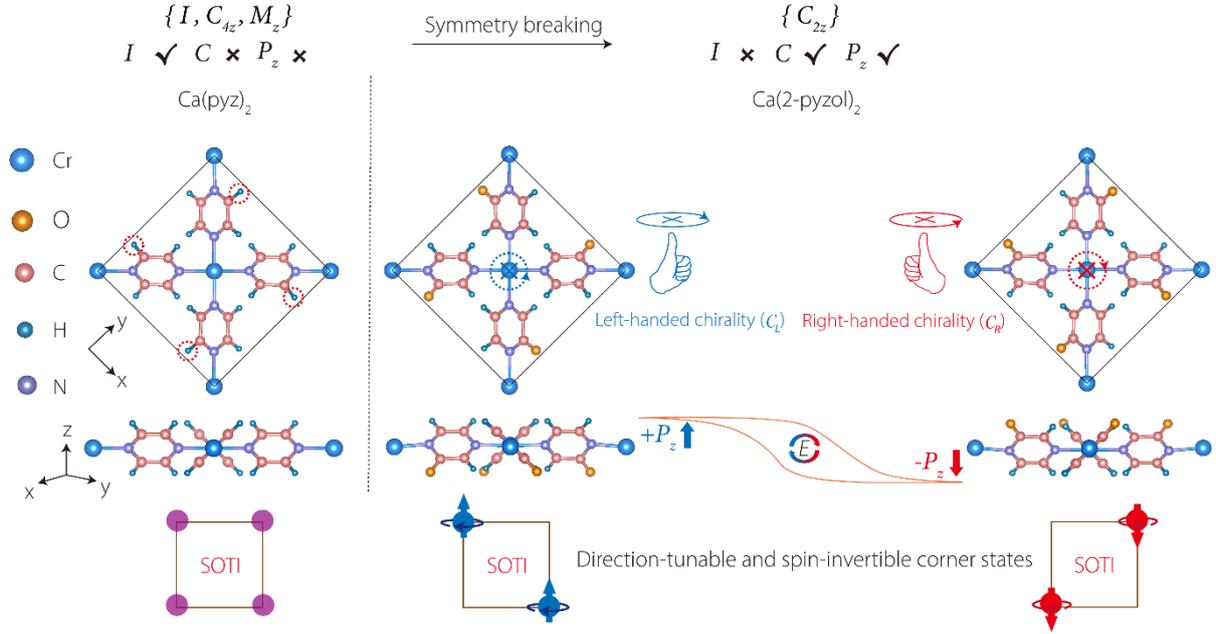

**Figure 1.** Structural transformation from Cr(pyz)$_2$ to Cr(2-pyzol)$_2$. The parent MOF Cr(pyz)$_2$ hosts $I$, lacks $P_z$ and $C$, and exhibits direction-independent corner states. By substituting four (out of 16) hydrogen atoms with oxygen (i.e., replacing pyz with 2-pyzol), Cr(2-pyzol)$_2$ emerges with $C_R$, which lacks $I$, hosts $-P_z$ and $C$, and exhibits direction-dependent corner states. The Cr(2-pyzol)$_2$ with $C_R$ and Cr(2-pyzol)$_2$ with left-handed chirality ($C_L$) correspond to two degenerate ferroelectric (DE-FE) states within the chiral ferroelectric MOF Cr(2-pyzol)$_2$. Here, $+P_z$ ($-P_z$) shows a state of total electric polarization that is outward (inward) perpendicular to the a-b lattice plane. The ferroelectric chirality switching significantly influences the corner states in Cr(2-pyzol)$_2$ from two perspectives: In real space, the direction of the corner states changes from (1,-1) to (1,1). In spin space, the magnetic spin of the corner states switches from up to down.

### 2.2. Ferroelectric Switchable Chirality

To identify the magnetic ground state of the Cr(2-pyzol)$_2$ chiral systems, eight distinct magnetic configurations are examined, including one ferromagnetic (FM) state, two antiferromagnetic (AFM) states, and five ferrimagnetic (FiM) states (see Figure S2, SI). The results indicate that the FiM1 state is the most stable, characterized by the antiparallel alignment of spins on metal nodes Cr with organic ligands 2-pyzol. The total and atomic magnetic moments for the FiM type-Cr(2-pyzol)$_2$ chiral systems are shown in Table S2 of the SI. Obviously, the magnetism in Cr(2-pyzol)$_2$ chiral systems primarily originates from the Cr atom, as is the case with Cr(pyz)$_2$[51,57]. The stability of Cr(2-pyzol)$_2$ is further confirmed through phonon dispersion curves, formation energy calculations, and ab initio molecular dynamics simulations (see Figure S3 and Figure S4 of the SI for details).

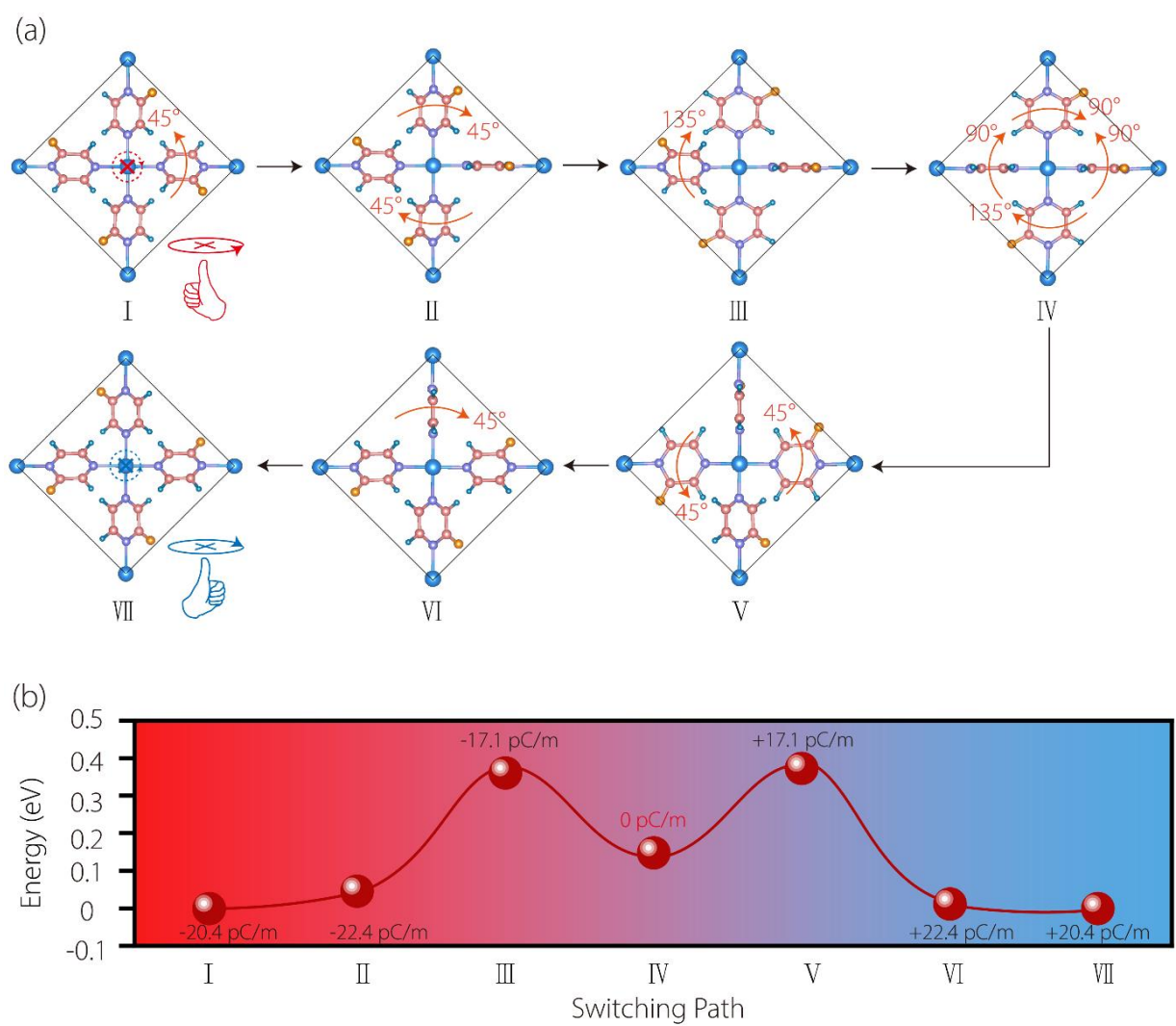

**Figure 2.** (a) A possible pathway and (b) energy barriers for the transition of Cr(2-pyzol)$_2$ from FE state I to FE state VII (from $C_R$ state to $C_L$ state).

Structurally, in the Cr(2-pyzol)$_2$ chiral system, the four 2-pyzol organic ligands surrounding the Cr atoms, arranged in either a clockwise or counterclockwise manner, act as four dipoles, so generating the electric polarization of chiral systems (as shown in Figure 1). Consequently,

Cr(2-pyzol)$_2$ with $C_R$ and $C_L$ can be regarded as two degenerate ferroelectric (DE-FE) states. The Cr(2-pyzol)$_2$ with $C_R$ state corresponds to a vortex state characterized by four clockwise dipoles arranged around the central Cr atom. Conversely, the Cr(2-pyzol)$_2$ with $C_L$ state corresponds to an antivortex state characterized by four counterclockwise dipoles arranged around the central Cr atom[49]. As shown in Figure 2, the polarization values of two DE-FE states (namely, states I and VII, corresponding to Cr(2-pyzol)$_2$ with $C_R$ and $C_L$) along the z-direction are -20.4 pC/m and 20.4 pC/m, respectively, which are comparable to those of reported 2D materials such as h-PbO (28 pC/m)[58], SnTe (22 pC/m)[59], and Co$_2$CF$_2$ (15.12 pC/m)[60].

Figure 2 depicts a potential transition pathway from state I ($C_R$ state) to state VII ($C_L$ state) in MOF Cr(2-pyzol)$_2$. Initially, in state I, the right dipole rotates counterclockwise by 45°, transitioning into state II. Then, the upper and lower dipoles rotate clockwise by 45°, shifting into state III. Next, the left dipole undergoes a 135° clockwise rotation, transitioning into state IV, an antiferroelectric (AFE) state with zero polarization (see Figure S5, SI). In this state, the upper and lower dipoles cancel each other in the ab-plane, while the left and right dipoles cancel along the z-direction, resulting in net-zero polarization. Subsequently, the upper and left dipoles rotate clockwise by 90°, the right dipole rotates counterclockwise by 90°, and the lower dipole rotates clockwise by 135°, yielding state V. Then, the left and right dipoles undergo a 45° counterclockwise rotation to reach state VI. Finally, a 45° clockwise rotation of the upper dipole, switching to state VII. The transition from state I to state III requires overcoming an energy barrier of 0.36 eV. Given that an appropriate ferroelectric transition barrier typically falls within 0.03–0.65 eV[48,49,61-65], this value suggests that Cr(2-pyzol)$_2$ can undergo electric-field-induced switching under practical conditions.

## 2.3. Magnetic SOTI Phase and Nontrivial Topological Invariants

In this section, we come to study the magnetic topological phases for the two DE-FE states of Cr(2-pyzol)$_2$ chiral systems, i.e., with $C_L$ and $C_R$. The spin-polarized band structures for two DE-FE states two states—FiM-type-Cr(2-pyzol)$_2$ with $C_R$ and $C_L$—are shown in Figure 3. Selecting FiM type-Cr(2-pyzol)$_2$ with $C_R$ (FE state I) as an example, the system is a magnetic semiconductor, with band gaps of 1.90 eV and 2.01 eV for spin-up and spin-down channels, respectively. For the Cr(2-pyzol)$_2$ with $C_L$ (FE state VII), this is a 180° reversal of magnetic spin occurs compared to Cr(2-pyzol)$_2$ with $C_R$, meaning the spin-up and spin-down bands in Cr(2-pyzol)$_2$ with $C_R$ are inverted in Cr(2-pyzol)$_2$ with $C_L$.

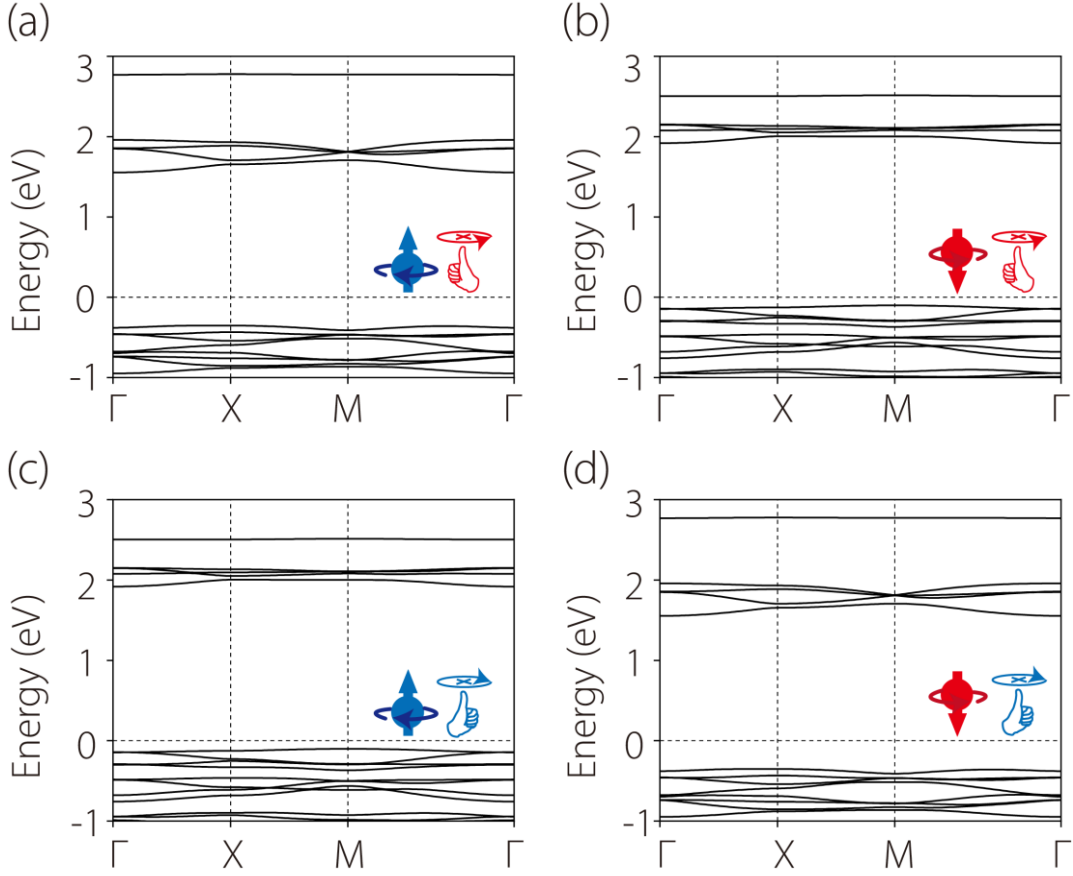

**Figure 3.** (a) and (b) Spin-up and spin-down band structures for Cr(2-pyzol)$_2$ with $C_R$ (FE state I), respectively. (c) and (d) Spin-up and spin-down band structures for Cr(2-pyzol)$_2$ with $C_L$ (FE state VII), respectively.

**Table 1.** Rotation topological invariants and quantized fractional corner charge for Cr(2-pyzol)$_2$ with $C_R$ (FE state I) in both spin channels.

| System | #$\Gamma_1^{(2)}$ | #$X_1^{(2)}$ | #$Y_1^{(2)}$ | #$M_1^{(2)}$ | $[X_1^{(2)}]$ | $[Y_1^{(2)}]$ | $[M_1^{(2)}]$ | $Q_c^{(2)}$ |
|---|---|---|---|---|---|---|---|---|
| Spin-up | 38 | 39 | 39 | 38 | 1 | 1 | 0 | $\frac{e}{2}$ |
| Spin-down | 42 | 43 | 43 | 42 | 1 | 1 | 0 | $\frac{e}{2}$ |

Without considering spin-orbit coupling, the spin-up and spin-down bands are decoupled, enabling us to treat each spin channel as an independent spinless system. In each spin channel of Cr(2-pyzol)$_2$, both $C_{2z}$ and $T$ exist, allowing for the calculation of the fractionally quantized corner charge to determine if each spin channel can be regarded as a 2D SOTI protected by $C_{2z}$. For a $C_{2z}$ invariant system, the rotation topological invariants are defined as[66]

$$\left[\Pi_p^{(2)}\right] \equiv \#\Pi_p^{(2)} - \#\Gamma_p^{(2)} \text{ for } p = 1, 2, \quad (1)$$

representing the difference in the number of states with $C_{2z}$-symmetric eigenvalues at the $\Pi_p^{(2)} = e^{\pi i(p-1)}$ points of Brillouin zone (BZ). To fully describe the $C_2$ symmetry-protected SOTI phases, the following independent rotation topological invariants should be considered[66]:

$$\chi^{(2)} = \{\left[X_1^{(2)}\right], \left[Y_1^{(2)}\right], \left[M_1^{(2)}\right]\}. \quad (2)$$

The quantized fractional corner charge $Q_c^{(2)}$, associated with $T$ and $C_{2z}$ symmetries, follows the relation[66]:

$$Q_c^{(2)} = \frac{e}{4}\left(-\left[X_1^{(2)}\right] - \left[Y_1^{(2)}\right] + \left[M_1^{(2)}\right]\right) \bmod e \quad (3)$$

, where $e$ is the charge of a free electron. Table 1 shows that the rotation topological invariants for Cr(2-pyzol)$_2$ with $C_R$ (FE state I) are $\chi^{(2)} = \{1,1,0\}$ for both spin channels, and the derived $Q_c^{(2)} = \frac{e}{2}$ for both spin channels, indicating that Cr(2-pyzol)$_2$ with $C_R$ is a magnetic SOTI[66,67]. Similarly, the Cr(2-pyzol)$_2$ with $C_L$ (FE state VII) is also a magnetic SOTI, as shown by the rotational topological invariants and $Q_c^{(2)}$ results in Table S3 of the SI.

## 2.4. Ferroelectric Chirality-Controlled Tunable Corner States

The 2D SOTIs will host unconventional boundary states that are localized at 0D corners of the nanodisk. To verify the corner states for the Cr(2-pyzol)$_2$ chiral systems in both spin channels, we construct a tight-binding model for a finite-size square-shaped nanodisk derived from the $C_2$-symmetric Cr(2-pyzol)$_2$ chiral systems with $C_R$ and $C_L$ (FE states I and VII). The energy spectra of the square-shaped nanodisk for both spin channels in Cr(2-pyzol)$_2$ with $C_R$ and $C_L$ are presented in Figure 4. For Cr(2-pyzol)$_2$ with $C_R$ (FE state I), four groups of twofold degenerate states (marked with red dots) are observed within the bulk and surface states (shown as black dots) in Figures 4a and 4b. Two groups of states, named A and B, belong to the spin-up channel, while the other two groups of states, labeled C and D, belong to the spin-down channel. When switching to Cr(2-pyzol)$_2$ with $C_L$ (FE state VII), the 180° inversal of magnetic spin compared to Cr(2-pyzol)$_2$ with $C_R$ leads to the inversion of the spin channels for states A-B and states C-D. Specifically, as seen in Figures 4c and 4d, the A and B states in Cr(2-pyzol)$_2$ with $C_L$ belong to the spin-down channel, while the C and D states belong to the spin-up channel.

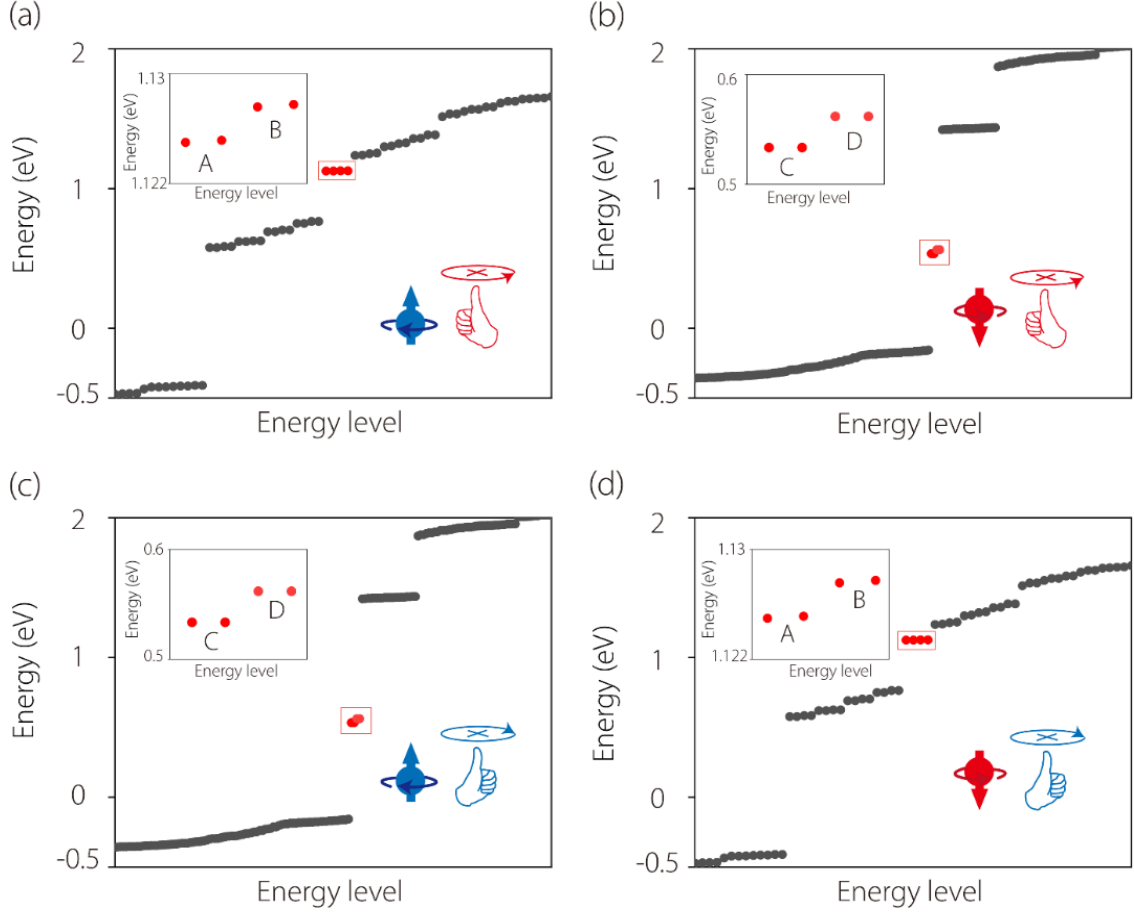

**Figure 4**. (a) and (b) The energy spectrums for Cr(2-pyzol)$_2$ with $C_R$ (FE state I) on a finite-size square-shaped nanodisk. (c) and (d) The energy spectrums for Cr(2-pyzol)$_2$ with $C_L$ (FE state VII) on a finite-size square-shaped nanodisk. (a) and (c) for the spin-up channel, and (b) and (d) for the spin-down channel, respectively. In (a) and (b) ((c) and (d)), four groups of twofold degenerate states (red dots; labeled A-D) are observed.

The spatial distributions of the four groups of states for Cr(2-pyzol)$_2$ with $C_R$ and $C_L$ (FE states I and VII) are plotted in Figure 5. These twofold degenerate states (A-D states) are localized corner states located at two of the four corners of the finite-size square-shaped nanodisk. Notably, due to the absence of $C_{4z}$ symmetry while maintaining $C_{2z}$ symmetry, the corner states are positioned along (1,1) or (1,-1) directions in real space. Therefore, one can conclude that the corner states in Cr(2-pyzol)$_2$ are direction-dependent, in contrast to the direction-independent corner states in Cr(pyz)$_2$, as reported by Wang e*t al.* [51].

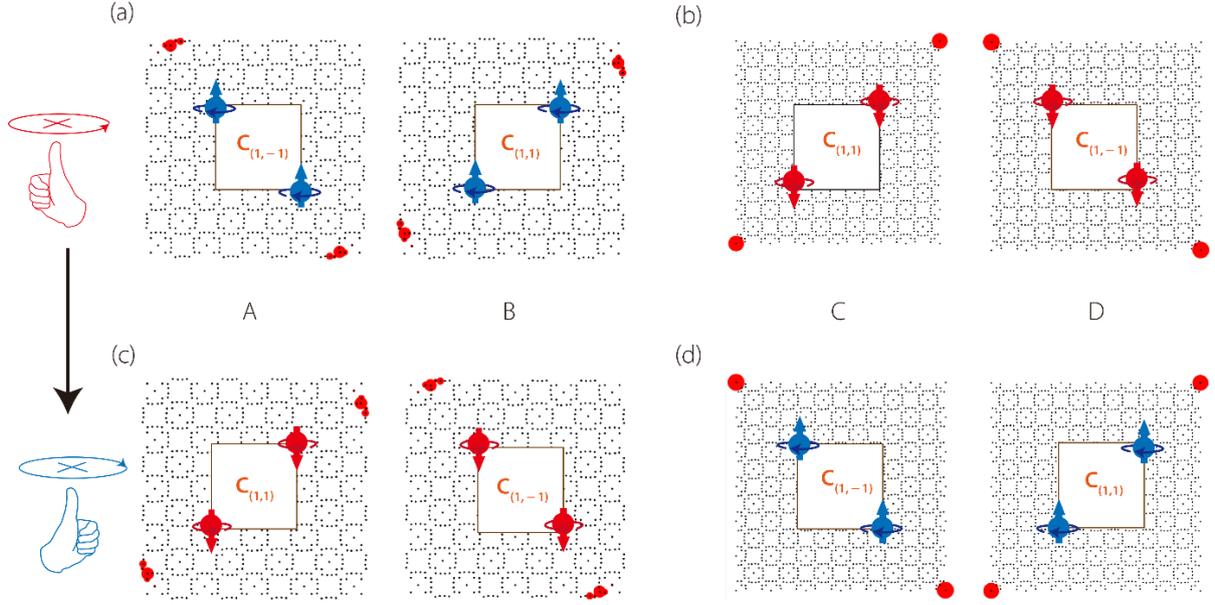

**Figure 5.** The spatial distribution of the four groups of corner states (A-D) for Cr(2-pyzol)$_2$ chiral systems with (a and b) $C_R$ (FE state I) and (c and d) $C_L$ (FE state VII). C$_{(1, -1)}$ and C$_{(1, 1)}$ represent the corner states along the (1, -1) and (1, 1) directions in real space, respectively. The blue upward arrow and red downward arrow indicate the corner states as spin-up and spin-down states in spin space, respectively.

Interestingly, switching the ferroelectric chirality from $C_R$ to $C_L$—i.e., changing the FE states from I to VII—allows the corner states in both spin space and real space to be tuned. For example, the state A in Cr(2-pyzol)$_2$ with $C_R$ is a localized corner state along the (1,-1) direction in real space and a spin-up state in spin space (see Figure 5a). Upon transitioning from $C_R$ to $C_L$, the A state becomes a corner state along the (1, 1) direction in real space and a spin-down state in spin space (refer to Figure 5c). The same behavior can also be observed for states B, C, and D. Therefore, it can be concluded that corner states in Cr(2-pyzol)$_2$ chiral systems are direction-tunable in real space and spin-invertible in spin space when the ferroelectric chirality is switched. Such electrically tunable spin configurations and topologically protected states present considerable possibilities for the development of next-generation topo-spintronic devices with multifunctional capabilities. This tunability facilitates the dynamic regulation of topological corner states with opposing spins, thereby paving the way for new memory architectures.

## 3. Conclusion

In conclusion, by selecting the 2D MOF-based magnetic SOTI Cr(pyz)$_2$ as an example, we break the $M_z$ and $C_{4z}$ symmetries and maintain the $C_{2z}$ symmetry. This is achieved by replacing pyz ligands with 2-pyzol ligands in Cr(pyz)$_2$, resulting in the Cr(2-pyzol)$_2$ MOF-based magnetic SOTI. In Cr(2-pyzol)$_2$, the corner states are direction-dependent and present in both spin-up and spin-down channels. The corner states are localized at two of the four corners of the square-shaped nanodisk, which contrasts with the direction-independent, spin-up corner states that are localized at all four corners of the square-shaped nanodisk in Cr(pyz)$_2$. Additionally, we demonstrate that ferroelectric chirality manifests in 2D Cr(2-pyzol)$_2$ MOF. The Cr(2-pyzol)$_2$ with $C_R$ and $C_L$ corresponds to two DE-FE states, which can be switched under an electric field, with a potential transition pathway for Cr(2-pyzol)$_2$ featuring an appropriate FE transition barrier of 0.36 eV. Remarkably, we demonstrate that the spin-polarized, direction-dependent corner states can be controlled via ferroelectric chirality. The direction of corner states in real space and the spin of corner states in spin space can both be simultaneously controlled by the switching of ferroelectric chirality. Our findings elucidate the emergence of ferroelectric chirality in 2D MOF-based magnetic SOTIs and offer a strategy to simultaneously control corner states in both real and spin spaces.

**Supporting Information**

Supporting Information is available from the Wiley Online Library or from the author.